\documentstyle[12pt,a4]{article}
\begin{document}
\newcommand{\ds}{\displaystyle}

\begin{center}

{\Large\bf Charge distribution in soft chiral pion bremsstrahlung.}

\vspace{2mm}

{\large I.V.~Andreev$^{1}$, M.~Biyajima$^{2}$,
V.A.~Nechitailo$^{1}$, N.~Suzuki$^{3}$}

\vspace{2mm}
{\it ${}^{1}$P.N.~Lebedev Physical Institute, Moscow, Russia}

{\it $^{2}$Department of Physics, Faculty of Liberal Arts,\\
Shinshu University, Matsumoto 390, Japan }

{\it $^{3}$Matsusho Gakuen Junior College, Matsumoto 390-12, Japan }

\end{center}

\begin{abstract}
Bremsstrahlung of soft chiral pions in high-energy processes is considered.
The distribution over number of neutral and charged pions is shown to obey
the inverse square root law. This law is argued to be generic for
multiple coherent production of soft chiral pions.
\end{abstract}

\section{Introduction}

Charge distribution of pions in multiple production processes
drew much attention recently. The growth of interest in this
subject is due to expectations to detect the disoriented chiral
condensate (DCC) formation in high energy collisions \cite{1}-\cite{6}.
The simplest picture of the process is given by "Baked Alaska
scenario" \cite{4}, where coherent pulses of semiclassical pion
field are emitted leading to anomalously large
fluctuations in the ratio of neutral to charge pions produced.
In particular, the probability to produce $n_o$ neutral pions
(for large total number of pions $n$) is given by inverse square
root formula,
\begin{equation}
w(n_0)\sim 1/\sqrt{n_0 n} ,
\label{1}
\end{equation}
being very flat and so quite different from usual binomial-like
distributions. This mechanism may be relevant for description of
"Centauro" (and "Anti-Centauro") type events found in cosmic ray
experiments, see \cite{8,9} and references therein, in which the number
of charged particles drastically exceeds the number of neutral ones
(or vice versa).

Now the problem arises -- to what extent the behaviour in Eq.~(\ref{1})
can be considered as a signature of DCC formation. Let us remind in
this connection that the distribution of the form of Eq.~(\ref{1})
was found long ago in a model of independent coherent
 pion production when
isotopic spin conservation was taken into account \cite{10,10b}.
In these and previous works \cite{11,12} the coherent production
of pions was taken for granted.More recently the topic of coherent
and squeezed states was discussed in the literature in the context
of DCC \cite{13}-\cite{18}.It was found that charge distribution
in squeezed states,having small isotopic spin,is also very broad \cite{18}.

In the present paper we consider a concrete mechanism
of pion production -- soft chiral pion bremsstrahlung
accompanying some basic high-energy
process and estimate charge distribution of the pions. The quantum charge
states of chiral pions emitted from simple vertices will be explicitly
calculated. It will be found that neutral pion number distribution
again has the form of Eq.~(\ref{1}). That is, such flat charge
distributions are typical for soft chiral pions and do not indicate
directly on DCC formation.

The soft chiral pion bremsstrahlung was studied many years ago \cite{19,20}.
Similarly to photons, soft pions are emitted from external lines of
diagrams representing the basic process (to be definite, we shall take
external particles to be spin 1/2 fermions (nucleons)). The complications
arising due to noncommutative pion-nucleon vertices and nonlinear pion-nucleon
coupling were shown to be mutually cancelled \cite{19}. Nonlinear pion-pion
coupling can be taken into account but its effect vanishes in the limit of
small pion momenta. Soft virtual pion exchange \cite{21} changes normalization
and does not influence the pion number distributions. The net result for
soft pion emission is given by substitution \cite{20}:
\begin{equation}
\psi_j \rightarrow \exp(-i\gamma_5\tau_i\phi_i/2)\psi_j,
\label{2}
\end{equation}
where $\psi_j$ is the fermion field for every incoming or outgoing
nucleon in the skeleton diagram of a basic process, $\phi_i=\pi_i/f_\pi$,
$\pi_i$ being the pion field, $f_\pi=93$ MeV is the pion decay constant.
The substitution in Eq.~(\ref{2}) can be deduced \cite{20,22} from
requirement for the effective lagrangian (S-matrix) of strongly interacting
fermions with accompanying soft pion emission to be chiral invariant
if it is isotopic invariant without these additional pions.

To simplify calculations we shall consider here the skeleton vertex
for two fermions, that is $\bar{\psi}\Gamma\psi$, where $\Gamma$ may
contain Dirac matrices $\gamma_{\mu}$ and (or) isotopic Pauli matrices
$\tau_i$. In the absence of isotopic matrices the
vector ($\Gamma=\gamma_{\mu}$) and axial ($\Gamma=\gamma_{\mu}\gamma_{5}$)
vertices are chiral invariant and do not produce the additional soft
pions we are considering here. Other vertices do allow the emission of
pions.

\section{Scalar vertex}

As the simplest example consider the scalar vertex $\bar{\psi}\psi$
($\Gamma$ is the identity matrix). Its chiral-invariant extension has the
form

\begin{equation}
V_s = \bar{\psi} \exp(-i\gamma_5\tau_i\phi_i)\psi(x).
\label{3}
\end{equation}

\noindent
It coincides formally with the modified nucleon mass term in the chiral
lagrangian. We neglect pion momenta and for the fields $\phi_i$ use
the decomposition

\begin{equation}
\phi_i\rightarrow \phi_{i}(0) = \phi_{i}^{+} + \phi_{i}^{-} =
\int d^3 k f(k)[a^{+}_{i}(k) + a^{-}_{i}(k)],
\label{4a}
\end{equation}

\noindent
where creation and annihilation operators $a^{+}_{i}(k), a^{-}_{i}(k)$
obey canonical commutation relations. In the free field approximation

\begin{equation}
f(k)=(2\pi)^{-3/2}(2k_0)^{-1/2}f^{-1}_{\pi}, \quad k_0\le k_{m}
\label{4b}
\end{equation}
where $k_{m}$ is an upper limit of pion softness.The most severe estimation
for this limiting momentum is the rho-meson mass ensuring applicability
of the chiral lagrangian technique though it may well appear to be
higher.In any case it does not exceed the momentum transfer $\Delta p$
in the vertex $\Gamma$,$k_{m}<\Delta p$.
To calculate the matrix elements of $\pi^{0}, \pi^{+}, \pi^{-}$
production

\begin{equation}
M_s=\langle n_{+},n_{-},n_{0}|\exp(-i\gamma_5\tau_i\phi_{i}(0))|0\rangle
\label{5a}
\end{equation}
it appears convenient to use the integral representation

\begin{equation}
\exp(-i\gamma_5\tau_i\phi_i) =
\frac{1}{4\pi} \int d\Omega\, e^{\ds ie_k\phi_k}
(1+ie_k\phi_k-i\gamma_5\tau_k\phi_k)
\label{5b}
\end{equation}
where $\vec{e}$ is the unit vector in three dimensions and integration
is performed over solid angles of the vector $\vec{e}$.

The total isotopic spin of the pions produced can be zero or one. Consider
the first case and decompose the pion fields into creation and
annihilation parts, $\phi = \phi^{+} + \phi^{-}$.
Using equations

\begin{eqnarray}
[\phi^{-}_{i},\phi^{+}_{i}]=c\delta_{ij}, \quad
e^{\ds ie_k\phi_k}|0\rangle = e^{\ds-c/2} e^{\ds ie_k\phi^{+}_{k}}|0\rangle
\label{6}
\\
c=\int d^3 k |f(k)|^2
\nonumber
\end{eqnarray}
we are led to consider a pion state of the form:

\begin{equation}
|\Phi_s\rangle=\left(1+\frac{d\ }{d\alpha}\right)_{\alpha=1}\frac{1}{4\pi}
\int d\Omega\, e^{\ds -\alpha^2c/2 + i\alpha e_k\phi^{+}_{k}}|0\rangle.
\label{7}
\end{equation}
The normalization factor $N_s$ of the state is

\begin{equation}
N_s=\langle\Phi_s|\Phi_s\rangle=
\frac{1}{2}\left[1-(4c-1)e^{\ds-2c}\right]
\label{8}
\end{equation}
and the average number of pions produced is

\begin{equation}
\langle n\rangle=\frac{1}{N_s}
\langle\Phi_s|\int d^3 k a^{+}_i(k) a^{-}_i(k)|\Phi_s\rangle=
c\frac{1+(4c-1)e^{-2c}}{1-(4c-1)e^{-2c}}.
\label{9a}
\end{equation}
If the average number of pions is large (it is the most interesting
case) then

\begin{equation}
\langle n\rangle \cong c, \quad c\gg 1.
\label{9b}
\end{equation}
To estimate it take the free field approximation (\ref{4b});
then

\begin{equation}
\langle n\rangle = \frac{1}{f^2_{\pi}}\int\frac{d^3 k}{(2\pi)^3 2k_0}
\cong  \frac{k_{m}^2}{8\pi^2 f^2_{\pi}}
\label{10a}
\end{equation}
that is, to produce many soft pions by the present mechanism one needs
momentum transfer in the vertex to be large enough

\begin{equation}
\Delta p >k_{m} >2\pi f^2_{\pi} \cong 0.6\ \mbox{\rm GeV}.
\label{10b}
\end{equation}

A prominent feature of the model is the distribution over number
of charged and neutral pions produced. It can be obtained from
matrix elements (\ref{5a}) and has the multiplicative form with
respect to the total number of pions $n=n_0+n_{c}$ and the number
of neutral pions $n_0$,

\begin{equation}
w(n,n_0) = w(n)w_n(n_0)
\label{11a}
\end{equation}
where

\begin{equation}
w_n(n_0) = \frac{1}{n+1} \frac{(n/2)!}{\Gamma(\frac{n+1}{2})}
\frac{\Gamma(\frac{n_0+1}{2})}{(n_0/2)!}
 \approx \frac{1}{\sqrt{n n_0}}
\label{11b}
\end{equation}
is the probability to produce $n_0$ neutral pions for the given
total number of pions ($\Gamma(n)$ is the Euler $\Gamma$-function,
$n_0$ and $n$ are even, $n_0\le n$), and

\begin{equation}
w(n) = \frac{(n-c+1)^2}{N_s}
\frac{e^{-c}c^n}{(n+1)!}
\label{11c}
\end{equation}
is the distribution over the total number of pions produced.

The distribution (\ref{11b}) over the number of neutral pions $n_0$
and corresponding distribution over the number of charged pions

\begin{equation}
w_n(n-n_{c}) \approx \frac{1}{\sqrt{n(n-n_{c})}}
\label{11d}
\end{equation}
are very broad. This means that the probabilities for events to have
almost all pions being charged or neutral are not negligible even for
$n\gg 1$. These distributions appear to be very similar to those
invented some time ago \cite{10,10b} for the explanation of Centauro-type
events.
The distribution
(\ref{11c}) over the total number of pions is Poisson-like (though with
an additional central dip at $n\sim\langle n\rangle$) and it is much
more narrow than (\ref{11b}),(\ref{11d}).

\section{Electromagnetic vertex}

As a case of immediate physical interest consider the soft pion emission
for electromagnetic scattering of strongly interaction fermions.
The skeleton vertex has now the form

\begin{equation}
V_0 = e \bar{\psi}\gamma_{\mu}Q\psi =
e \bar{\psi}\gamma_{\mu}\frac{\tau_3+N_B}{2}\psi
\label{12}
\end{equation}
where the baryon number $N_B = 1$ for nucleons and $Q$ is electric
charge in units of $e$. Chiral extension of the vertex is taken as

\begin{equation}
V = e \bar{\psi}e^{\ds-i\gamma_5\tau_k\phi_k/2}\gamma_{\mu}
\frac{\tau_3+N_B}{2}e^{\ds-i\gamma_5\tau_k\phi_k/2}\psi
\label{13}
\end{equation}
leading to the substitution

\begin{equation}
\gamma_{\mu}\tau_3\rightarrow
\gamma_{\mu}\tau_3 +
\gamma_{\mu}\gamma_{5}\varepsilon_{ij3}\phi_i\tau_j\frac{\sin\phi}{\phi}
-\gamma_{\mu}(\tau_3\phi^2-\phi_3\tau_k\phi_k)
\frac{1-\cos\phi}{\phi^2}, \quad
\phi=\sqrt{\phi^2_k}
\label{14}
\end{equation}

We consider here diagonal transitions. Then the pion state has the form

\begin{eqnarray}
|\Phi_e\rangle &=&
(\phi^2_1+\phi^2_2)(1-\cos\phi)/\phi^2|0\rangle =
\nonumber
\\
&=&\frac{1}{4\pi}\int\limits_{|\vec{x}|\le1}
\frac{d^3 x}{|\vec{x}|}
\left( -\frac{\partial^2\ }{\partial x^2_1}
       -\frac{\partial^2\ }{\partial x^2_2}  \right)
e^{\ds i x_k\phi_k}|0\rangle
\label{15}
\end{eqnarray}
and its isotopic spin is equal to zero. The normalization
factor is

\begin{eqnarray}
N_e=\langle\Phi_e|\Phi_e\rangle =
\frac{4}{5}+\frac{16}{15}(c-1)e^{-c/2}
-\frac{4}{15}(4c-1)e^{-2c}
\label{16}
\\
N_e=\left\{
\begin{array}{l}
2c^2    \quad \mbox{\rm for}  \quad       c\ll 1\\
4/5     \quad \mbox{\rm for}  \quad       c\gg 1
\end{array}
\right.
\nonumber
\end{eqnarray}
where $c$ is given by Eq.~(\ref{6}) and the average number of
pions is

\begin{eqnarray}
\langle n\rangle =
\frac{4}{15N_e}\left[c+3-4e^{-c/2}
+(4c^2-c+1)e^{-2c}\right]
\label{17a}
\\
\langle n\rangle =
\left\{
\begin{array}{l}
1           \ \quad \mbox{\rm for}  \quad       c\ll 1\\
\frac{1}{3}c  \quad \mbox{\rm for}  \quad       c\gg 1
\end{array}
\right.
\label{17b}
\end{eqnarray}

As in the previous case, the most interesting characteristic is the
distribution over the number of pions of different charges.
Considering the state (\ref{15}) one has to calculate the matrix
elements

\begin{eqnarray}
M_e & = & N_e^{-1/2}\langle n_{+},n_{-},n_{0}|\Phi_e\rangle =
\nonumber
\\
&=&N_e^{-1/2}\frac{1}{4\pi}
\int\limits_{|\vec{x}|\le1}
\frac{d^3 x}{|\vec{x}|}
\left( -\frac{\partial^2\ }{\partial x^2_1}
       -\frac{\partial^2\ }{\partial x^2_2}  \right)
\langle 0|a_{i_1}(q_1)\dots a_{i_n}(q_n) e^{\ds i x_k\phi_k}|0\rangle
\nonumber
\\
&=&K\int\frac{d^3 x}{2\pi|\vec{x}|}
\frac{\partial\ }{\partial x_{+}}
\frac{\partial\ }{\partial x_{-}}
\left[ e^{\ds-cx^2/2}x^{n_{+}}_{+}x^{n_{-}}_{-}x^{n_{0}}_{3}
\right]
\label{18}
\end{eqnarray}
where

\begin{equation}
K=\frac{-i^n}{N_e^{1/2}}\prod^{n}_{j=1}f(q_j), \quad
\prod_j\int dq_j|K|^2=\frac{c^n}{N_e^{1/2}},  \quad
x_{\pm} =\frac{1}{\sqrt{2}}(x_1\pm i x_2).
\label{19}
\end{equation}
The matrix element for transition to the state, which contains $n_0$
neutral pions and $n_c$ charged pions can be represented as the
sum of two terms:

\begin{eqnarray}
M_e  =
\frac{K}{c^{n/2}}
\frac{2^{n_0/2}\Gamma(\frac{n_0+1}{2})(\frac{n_c}{2})!}{\Gamma(\frac{n+3}{2})}
\nonumber
\\
\left[
\frac{(n-3n_0)}{2n(n+3)}\gamma\left(\frac{n}{2}+1,\frac{c}{2}\right) +
\left(\frac{n_c(n+1)}{2n} - \frac{c(n_c+2)}{2(n+3)}\right)
\left(\frac{c}{2}\right)^{n/2}
 e^{-c/2}
\right]
\label{20}
\end{eqnarray}
where $n_c$ and $n_0$ are even, $n=n_0+n_c$, $n\not=0$ and

$$
\gamma\left(\frac{n}{2}+1,\frac{c}{2}\right)=
\int\limits^{c/2}_{0}du\, u^{n/2} e^{-u}
$$

Consider the most interesting situation, when the average number of
pions in the final state in Eq.~(\ref{18}) is large, $c\gg 1$.
Then the first term in square brackets in Eq.~(\ref{20})
dominates for small numbers of pions, $n\sim 1$, and the second term
dominates for large $n\sim c$,
their interference being small. Therefore the second term gives the
distribution over the number of neutral and charged pions in high
multiplicity events.  It reads:

\begin{eqnarray}
w_2(n,n_0)  \cong
\frac{3(c-n_0)^2}{2N_ec^{2}\sqrt{2c}}
\frac{\Gamma(\frac{n_0+1}{2})}{\left(\frac{n_0}{2}\right)!}
        w_2(n)
\label{21a}
\\
w_2(n)  \cong
\frac{2}{3c\sqrt{2\pi c}}(n-c)^2
\exp\left(-\frac{(n-c)^2}{2c}\right)
\label{21b}
\end{eqnarray}
where $w_2(n)$ is the probability to find $n=n_0+n_c$ pions, $n$
and $n_0\le n$ are even, $N_e\cong4/5$.

The distribution over the number of neutral (or charged) pions in
Eq.~(\ref{21a}) is again very broad ensuring a sizable number of
events, in which almost all pions are neutral (or charged).
The distribution $w_2(n)$ over the total number of pions is
again narrow and in fact coincides with Eq.~(\ref{11c}) for
$c\gg 1$, $n\gg1$ up to a normalization factor. The total probability
of high multiplicity events is

\begin{equation}
\sum_{n=2k} w_2(n)=\frac{1}{3}
\label{22}
\end{equation}
just corresponding to factor $1/3$ in Eq.~(\ref{17b}).

Small multiplicity events are given by the first term in square brackets
in Eq.~(\ref{20}). For large  average multiplicities $c\gg 1$
the corresponding probability is

\begin{eqnarray}
w_1(n\not=0,n_0)  \cong
\frac{1}{N_e}
\frac{\sqrt{\pi}\Gamma(\frac{n_0+1}{2})}{\left(\frac{n_0}{2}\right)!}
\frac{(n-3n_0)^2}{4n^2(n+3)^2}
\frac{\Gamma^2(\frac{n}{2}+1)}{\Gamma^2(\frac{n}{2}+\frac{3}{2})}
\label{23a}
\\
w_1(0,0) \cong 5/9
\label{23b}
\end{eqnarray}
where $n$ and $n_0$ are even and $n_0\le n$. Performing the sum over the
number of neutral pions in Eq.~(\ref{23a}) we obtain the distribution over
the total number of pions in low multiplicity events:

\begin{eqnarray}
w_1(n) & = &
\frac{1}{2n(n+3)}
\frac{\sqrt{\pi}\Gamma\left(\frac{n}{2}+1\right)}
{\Gamma\left(\frac{n}{2}+\frac{3}{2}\right)},
\quad n\not=0,\  c \gg 1
\label{24}
\\
\sum_{n=2k\not=0} w_1(n) & = & \frac{1}{9}
\label{25}
\end{eqnarray}
According to Eq.~(\ref{24}), the average number of $\pi$ pairs in low
multiplicity events is very small, equal to $1/3$ on the average,

\begin{equation}
\sum_{n=2k} \frac{n}{2}w_1(n) = \frac{1}{3}.
\label{26}
\end{equation}

\section{Discussion and conclusions}

Two examples of soft chiral pion emission considered above show very
broad distributions over the number of charged and neutral pions in
high multiplicity events. The neutral pion distributions in both
cases are given essentially by the function

\begin{eqnarray}
w(n_0) \cong
\frac{1}{\sqrt{2\langle n\rangle}}
\frac{\Gamma\left(\frac{n_0}{2}+\frac{1}{2}\right)}
{\Gamma\left(\frac{n_0}{2}+1\right)},
\qquad (n_0\ \mbox{\rm even})
\label{30a}
\\
w(n_0) \cong
\frac{1}{\sqrt{\langle n\rangle n_0}}
\qquad \mbox{\rm for}\ n_0\gg1, \ n_0\ \mbox{\rm even}
\label{30b}
\end{eqnarray}
which is similar to Eq.~(\ref{1}). This distribution gives the
sizable probability to find events with a small number of neutral
pions. For example, the probability to find no neutral pions is

\begin{equation}
w(0) =\sqrt{\pi/2\langle n\rangle}
\label{30c}
\end{equation}
amounting here to more than 10\%
for $\langle n\rangle=100$.
As the distribution over the total number of pions $n$ is rather
narrow for $\langle n\rangle \gg 1$, the charged pion distribution
is given mainly by the same function with substitution $n_0=n-n_{c}$.

The conditions for such broad distributions to appear in high
multiplicity events are small isotopic spins of the pion system
and many particle matrix elements symmetric with respect to pion
momenta, thus ensuring a constructive interference. In other words,
the pion emission must be coherent\footnote{
In its simplest phenomenological version the corresponding pion
state is the eigenstate of the annihilation operator of isoscalar
pion pair and it can be considered as a coherent state for an isoscalar
pair, see (\cite{10b})}.
This can be seen already from an early paper by A.~Pais \cite{15}
and was explicitly demonstrated more recently in paper \cite{16}.
Both of these conditions are fulfilled in our chiral model examples.

The bremsstrahlung spectrum of pseudoscalar pions has the form
$dn\sim kdk$ (contrary to photon spectrum $dk/k$) and so very small
momenta $k$ are inefficient for this mechanism.It was necessary
(as in Bloch-Nordsiek model) to introduce an upper limit of pion softness,
$k<k_{m}$ and the total number of pions produced by this mechanism
is proportional to $k_{m}^2$.The value of $k_m$ is not quite definite
(the most severe possible estimation is around rho-meson mass)
but it does not exceed the momentum transfer $\Delta p$ in the baryonic vertex
$\Gamma$.Anyhow it is clear that the presence of large baryonic momentum
transfer $\Delta p$ (and so the presence of high $p_{T}$ baryons)
is highly favourable for copious production of pions by the present
mechanism.At the same time the soft pions are expected to be present
in lower $p_{T}$ region.In the last region the pion spectrum of the form
$kdk$ by itself can be used for identification of the process of pion
bremsstrahlung.It can be seen in future experiments when it will be possible
to look at narrow windows of $p_T$.

One can attempt to apply this mechanism for a description of "Centauro"
and "Anti-Centauro" events in cosmic rays. The average transverse
momentum of particles in these events is just very high, three to
six times the value typical for hadronic processes, see Ref. \cite{9},
where the compilation of exotic events in cosmic rays is given.

In conclusion, it thus appears that inverse square root
distributions over number of neutral and charged pions are of
very general nature being characteristic for coherent soft
pion radiation.

\bigskip
{\Large\bf\noindent Acknowledgments}
\bigskip

This work was supported in part by the JSPS Program on Japan-FSU
Scientists Collaboration. I.A. and V.N. were also supported by
Russian Fund for Fundamental Research,grant 96-02-16210a.

\end{document}